\documentstyle[12pt,epsfig]{article}

 \title{Nanotube diameter optimal for
 channeling of high-energy particle beam}

\author{V.M. Biryukov$^1$, S. Bellucci$^2$ \\
1 Institute for High Energy Physics, Protvino, Russia \\
2 INFN - Laboratori Nazionali di Frascati, 00044 Frascati, Italy}
\date{}
\begin{document}
\normalsize
\baselineskip=23pt
\maketitle

\abstract{
\normalsize
\baselineskip=23pt
Channeling of particle beam in straight and bent single-wall nanotubes has been studied in computer
simulations. We have found that the nanotubes should be sufficiently narrow in order to steer efficiently
the particle beams, with preferred diameter in the order of 0.5-2 nm. Wider nanotubes, e.g. 10-50 nm,
appear rather useless for channeling purpose because of high sensitivity of channeling to nanotube
curvature. We have compared bent nanotubes with bent crystals as elements of beam steering technique,
and found that narrow nanotubes have an efficiency of beam bending similar to that of crystals.
}
\pagebreak

\section{
Introduction}

Bent crystals have been efficiently used for channeling of particle beams at accelerators [1] in the energy range
spanning six decades, from 3 MeV [2] to 900 GeV [3]. Currently, bent crystals are largely used for extraction of 70-
GeV protons at IHEP (Protvino) with efficiency reaching 85\% at intensity well over 10$^{12}$ particle per second,
steered by silicon crystal of just 2 mm in length [4]. A bent crystal (5 mm Si) is installed into Yellow ring of the
Relativistic Heavy Ion Collider where it channels Au ions and polarised protons of 100-250 GeV/u as a part of the
collimation system [5].

Since 1991 [6], here has been a lot of study on carbon nanotubes to understand their formation and properties. Carbon
nanotubes stick out in the field of nanostructures, owing to their exceptional mechanical, capillarity, electronic
transport and superconducting properties [7-9]. They are cylindrical molecules with a diameter of order 1 nm and a
length of many microns [10]. They are made of carbon atoms and can be thought of as a graphene sheet rolled around a
cylinder [11]. Creating efficient channeling structures - from single crystals to epitaxial layers [2] to nanotubes  - might
have a significant effect onto the accelerator world. It is well known that nanotubes can be manufactured of different
diameter - from a fraction of nm up to microns, of different length - from tens of microns up to millimeters, of different
material - usually carbon but also others [12]. This makes nanotubes a very interesting object for channeling research.

It has been shown in refs. [13-15] that  in a carbon nanotube the channeled particles are confined in a potential well
$U(\rho,\phi)$ with the depth $U_0$ of about 60 eV. The critical angle for channeling is then
$\theta_c=(2U_0/pv)^{1/2}$, $pv$ is the particle momentum times velocity.
In a nanotube bent with radius $R$, similarly to bent crystals [1], an
effective potential taking into account a centrifugal term $pvx/R$ is introduced [15]:
$U_{eff}(\rho,\phi)=U(\rho,\phi)+pvx/R$, where $x=\rho \cos(\phi)$ is the
coordinate in the
direction of bending. As shown in [15],  a carbon nanotube of  1.4 nm
diameter has a significant effective potential well $U_{eff}$ even for bendings equivalent to $\simeq$300 Tesla or
$pv/R\ge$ 1 GeV/cm, i.e. bent nanotubes can be comparable to (110) and (111) channels in silicon crystal commonly
used in crystal channeling applications.

The present work is devoted to Monte Carlo simulation of particle channeling in bent single-wall nanotubes, aimed at
finding how useful are the nanotubes for channeling of positively-charged particle beams, what kind of nanotubes are
efficient for this job, and how nanotubes compare with crystals in this regard.

Two mechanisms of particle transfer from channeled to random states are well known for crystals: scattering on
electrons and nuclei and curvature of the channel [1]. Typical nanotube is on the order of 0.1 mm in length at present.
For such a short channel, the scattering on electrons within the bulk of the tube is quite small. However, a curvature of
the tube could quickly (in less than one oscillation) bring much of the channeled particles out of the potential well or
into close collisions with the nuclei of the nanotube walls.

\section{Simulation model}

In order to assess the general properties of channeling in nanotubes, we average the potential $U(\rho,\phi,z)$
of a straight nanotube over the longitudinal coordinate $z$ and angle $\phi$ to obtain a potential $U(\rho)$ with
cylinder symmetry.
As known from crystal channeling, the averaging over $z$ is well justified as a collision of a particle with a nanotube wall
under a glancing angle does involve many thousand atoms along the particle trajectory. For the same reason, the
averaging over $\phi$ is equally justified if the nanotube has arbitrary helicity [15] as defined by nanotube indices (m,
n). In the special cases of zigzag (m=0) or armchair (m=n) nanotubes, the wall consists of atomic rows parallel to the
nanotube axis; the nanotube potential is then defined by the sum of potentials of the rows. However, even in these cases
the averaged potential $U(\rho)$ is significantly different from $U(\rho,\phi)$ only in rather close vicinity
from the wall, $\le$0.1 nm [14,15], where scattering on the wall nuclei is also important. Further on in the paper we
apply only the averaged potential $U(\rho)$ for a straight nanotube.

We use so-called standard potential introduced by Lindhard [16]. When averaged over ($\phi, z$), it is described by
[13]:
\begin{equation}
U(\rho )=\frac{4NZ_1Z_2e^2}{3a}\ln
(\frac{r^2+\rho^2+3a_S^2+\sqrt{(r^2+\rho^2+3a_S^2)^2-4r^2\rho^2}}{2r^2})
 \end{equation}
Here $Z_1e, Z_2e$ are the charges of the incident particle and the nanotube nuclei respectively, $N$ is the number of
elementary periods along the tube perimeter, $a$=0.142 nm is the carbon bond length;
the nanotube radius is $ r=Na\sqrt{3}/2\pi$ .

The screening distance $a_S$ is related to the Bohr radius $a_B$  by
\begin{equation}
a_S=\frac{a_B}{2} (\frac{3\pi}{4(Z_1^{1/2}+Z_2^{1/2})})^{2/3}
\end{equation}
In a tube bent along the $x$ direction, the motion of a particle is described by the equations
\begin{equation}
pv\frac{d^2x}{dz^2}+\frac{dU(\rho)}{dx}+\frac{pv}{R(z)}=0
 \end{equation}
\begin{equation}
pv\frac{d^2y}{dz^2}+\frac{dU(\rho)}{dy}=0
 \end{equation}
where $\rho^2=x^2+y^2$. This takes into account only the nanotube potential and the centrifugal potential.
Any particle within close distance, order of $a_S$, from the wall (where density of the nuclei is significant) is also
strongly affected by the nuclear scattering.

\section{Optimisation of nanotube}

We did look in simulations how channeled particles survive in nanotubes of different curvatures and diameters.
Figure 1 shows three examples of trajectories for 1-GeV protons channeled in carbon nanotubes of 1.1, 2.2, and 4.4 nm
diameter, bent with the same curvature radius 1 cm. The range of transverse coordinate $x$ where channeled particles are
localised in the tube is about 0.5 nm in each of the three cases. Taking into account different diameters of the tubes,
these examples suggest that whereas the narrow tube is mostly filled with channeled particles, the wider tubes are
nearly empty. The centrifugal force kills channeling in most of the wide-tube cross-section where there is no field to
guide the particles.

To make a nanotube an efficient channeling structure, we must optimise the nanotube diameter with respect to the
curvature of the nanotube. It is rather easy to understand the match between the diameter and the curvature. In the time
needed for channeled particle to cross the nanotube (which is in proportion to the diameter), the angular orientation of
the bent tube should not change by more than the critical angle $\theta_c$, otherwise --
if particle comes to the wall
with an angle greater than $\theta_c$ -- it will cross the wall and become dechanneled.

Figure 2 gives another simple explanation. The top of the figure shows the potential well $U(\rho)$ for tubes of
different diameter $d$; the depth $U_0$  of each well is the same. The bottom of the figure illustrates the effective
potential $U_{eff}(x)=U(x)+ pvx/R$ of the same tubes bent with radius $R$ for particles of momentum $p$. Only minor part
of the wide well $U_{eff}$ can confine the channeled particles. If the change in the centrifugal potential $pv/R$ over the tube
diameter $d$ is much greater than $U_0$, certainly the tube is inefficient. So a rough estimate for the optimal diameter
$d_{opt}$  will come from the relation $(pv/R)d \simeq U_0$, that is $d_{opt}\simeq RU_0/pv$.

We can rewrite it using an equivalent magnetic field $H_{eq}$ corresponding to $pv/R$  ratio:
$$d_{opt}\simeq 1.8 {\rm nm} \frac{\rm 100 Tesla}{H_{eq}}$$
and substituting $U_0\simeq $ 60 eV  for the carbon nanotube. The typical range for $H_{eq}$   as used in bent
crystal channeling is 30-300 Tesla (the respective range of  pv/R is 0.1-1.0 GeV/cm),
and can be as much as $\simeq$1000 Tesla with
silicon (and 2-3 times higher with Ge and W). For nanotubes working in the same $H_{eq}$ range,
30-300 Tesla, the diameter should be in the range 0.6-6 nm. The stronger field we want to use for beam steering, the
thinner channel (nanotube) should be.
Interestingly, the range of 0.6-6 nm is just about the typical range of easily manufactured carbon nanotubes.

We also have to conclude that wider nanotubes, like $\simeq$60 nm, are good only for weak equivalent fields of
$\simeq$3 Tesla. Any substantial curvature of a wide nanotube would destroy most of channeling in it. Therefore the
interest to wide channeling structures motivated by an expected lack of scattering inside them is quite limited now (if
not canceled at all) by considerations of bending dechanneling.

The equation for $d_{opt}$  can be written in general form for nanotubes of other material, with the use of Lindhard
potential:
\begin{equation}
d_{opt}\simeq \frac{8\pi Z_1Z_2e^2a_S}{3a^2}\frac{R}{pv}
 \end{equation}

This short analysis suggests that, in order to compete with crystals at beam steering, the nanotube should be as narrow
as order of 1 nm.

\section{Results of simulation}

Following the above analysis, we have studied the efficiency of bent nanotube channeling in Monte Carlo simulations.
A parallel beam of 1 GeV protons was entering a carbon nanotube, where protons were tracked over 50$\mu$m.
Depending on the nanotube diameter, curvature and starting coordinates, every proton did 5 to 20 oscillations over that
distance, typically. Multiple scattering was not included, so we did evaluate only the effects of bending dechanneling.
Figure 3 shows an example of the angular distribution of particles downstream of the 50-$\mu$m long nanotube of
1.1 nm diameter, bent 4 mrad, shown in the direction of bending. Similarly to pictures of bent crystal channeling, there
is clear separation of channeled and nonchanneled peaks, with some particles lost (dechanneled along the tube) between
them.

We did such a simulation for variety of nanotubes with different characteristics. Figure 4 shows the number of protons
channeled through 50 $\mu$m as a function of the nanotube curvature $pv/R$ for tubes of different diameter. For
comparison, also shown is the same function for Si(110) crystal (from ref.[1]).

The channel length of 50 $\mu$m, with bending of 1 GeV/cm, gives the 1-GeV protons a deflection of 5 mrad -
sufficient for many accelerator applications like extraction [1,3-5]. One can see from Fig.4 that nanotubes as narrow as
1 nm are comparable to silicon crystals in beam bending. However, nanotubes of $\ge$2 nm diameter already show
poor behaviour in channeling for significant bending. The wider nanotube is, the weaker channeling properties it
demonstrates, in agreement with the discussion of previous section.

\section{Summary}

As shown in computer simulations, nanotubes can channel particle beams with efficiency similar to that of crystal
channeling. Bending dechanneling, present if nanotube has a curvature, is found to be a very significant factor in
nanotube channeling. It sets the nanotube diameter optimal for channeling to be in the range of about 0.5-2 nm. Much
wider nanotubes are of little use for channeling purpose because of high sensitivity of channeling to curvature. Narrow
nanotubes, with diameter on the order of 1 nm, can be a basis for an efficient technique of beam steering at particle
accelerators.

\section*{Acknowledgements}

This work was partially supported by INFN - Gruppo V, as NANO experiment, and by INTAS-CERN Grant No. 132-
2000 and RFBR Grant No. 01-02-16229.
\vspace{1cm}


{\large\bf References}
\vspace{5mm}

[1]	V.M. Biryukov, Yu.A. Chesnokov and V.I. Kotov, "Crystal Channeling and its Application at High Energy
	Accelerators" (Springer, Berlin, 1997)

[2]	M.B.H.Breese, Nucl. Instr. and Meth. B 132, 540 (1997)

[3]	R.A.Carrigan, Jr., et al. Phys. Rev. ST Accel. Beams AB 1, 022801 (1998)

[4]	A.G. Afonin, et al. Phys. Rev. Lett. 87, 094802 (2001)

[5]	R.P. Fliller III, et al., contributed to EPAC 2002 Proceedings (Paris)

[6]	S. Iijima, Nature 354, 56 (1991) Appl. Phys. Lett. 80, 2973 (2002)

[7]	M. Bockrath, et al., Nature 397, 598 (1999); Z. Yao, et al., Nature 402, 273 (1999)

[8]	S. Bellucci and J. Gonzalez, Eur. Phys. J. B 18, 3 (2000); ibid. Phys. Rev. B 64, 201106 (2001) (Rapid Comm.)

[9]	A. Yu. Kasumov, et al., Science 284,1508 (1999); M. Kociak, et al., Phys. Rev. Lett. 86, 2416 (2001)

[10]	R. Saito, G. Dresselhaus, M. S. Dresselhaus, "Physical Properties of Carbon Nanotubes" (Imperial College Press,
	London, 1998)

[11]	T.W. Ebbesen, Phys. Today, 49, 26 (1996)

[12]	Z.Y. Wu, et al., Appl. Phys. Lett. 80, 2973 (2002)

[13]	V.V.Klimov and V.S.Letokhov, Phys. Lett. A 222, 424 (1996)

[14]	L.G.Gevorgian, K.A.Ispirian, K.A.Ispirian. JETP Lett. 66, 322 (1997)

[15]	N.K.Zhevago and V.I.Glebov, Phys. Lett. A 250, 360 (1998)

[16]	J. Lindhard, K. Dan. Viddensk. Selsk. Mat. Phys. Medd. 34, 1 (1965)

\pagebreak

{\large\bf Figure captions}
\vspace{1cm}

{\bf Figure 1}

Trajectories of 1-GeV protons channeled in carbon nanotubes 
of 1.1 (a), 2.2 (b), and 4.4 (c) nm diameter, bent with the same
curvature radius 1 cm.

{\bf Figure 2}

Schematic potential well (a, b) U(x) for tubes of different diameter, and effective potential
 (c, d) $U_{eff}(x)= U(x)+ pvx/R$ of these tubes bent with some radius $R$. 
 The dotted line marks the top of the well confining
the channeled particles.

{\bf Figure 3}

The angular distribution of particles downstream of the bent nanotube, 
in the direction of bending.

{\bf Figure 4}

The number of channeled protons as a function of the nanotube curvature 
$pv/R$ for tubes of different diameter (bottom
up: 11, 2.2, and 1.1 nm). 
For comparison, also shown (top curve) is the same function for Si(110) crystal.

\pagebreak

 \begin{figure}[htb]
\begin{center}
\parbox[c]{10.5cm}{\epsfig{file=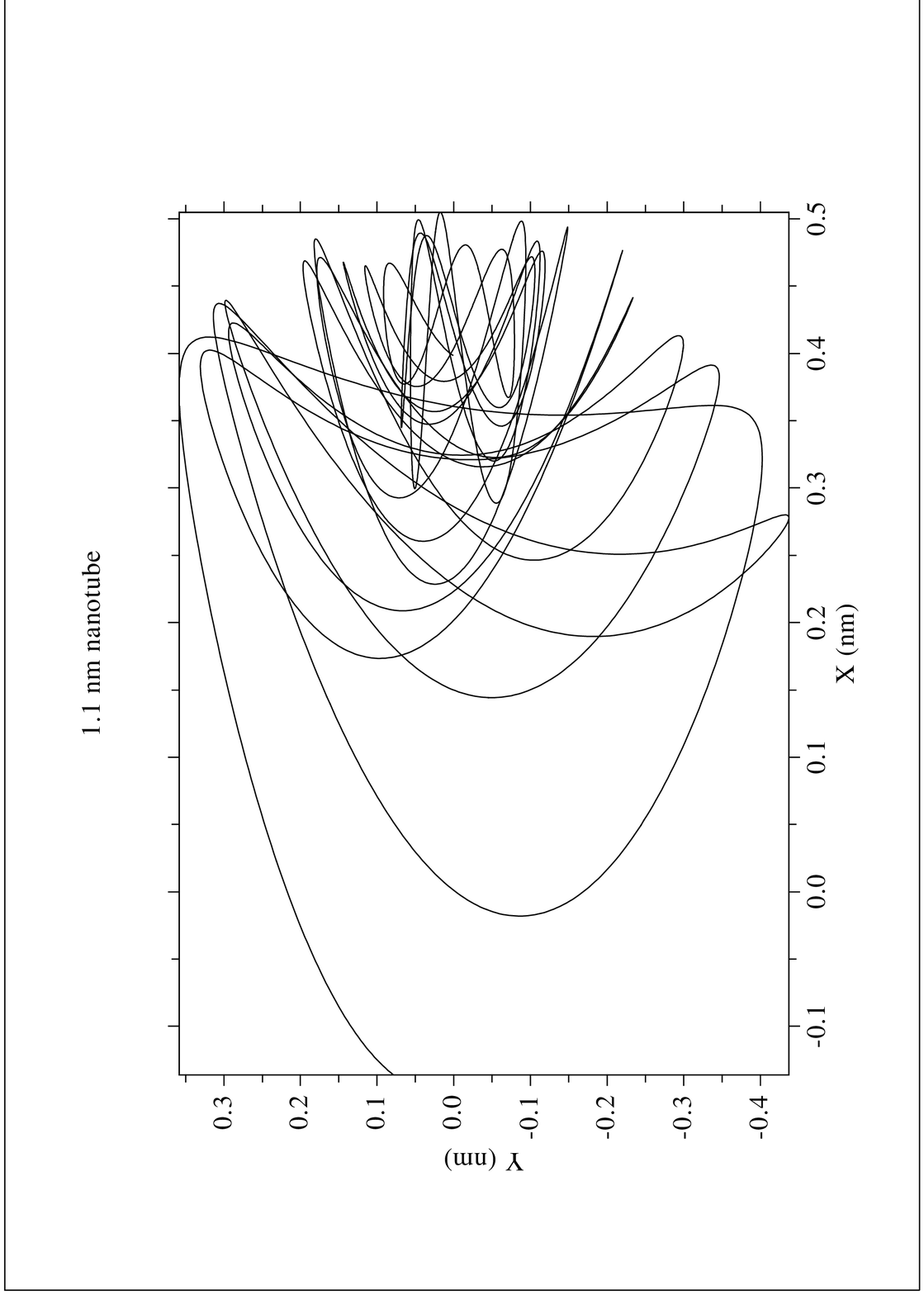,width=9 cm }}
\end{center}
        \caption{(a)
}
        \end{figure}

\pagebreak

\setcounter{figure}{0}
 \begin{figure}[htb]
\begin{center}
\parbox[c]{10.5cm}{\epsfig{file=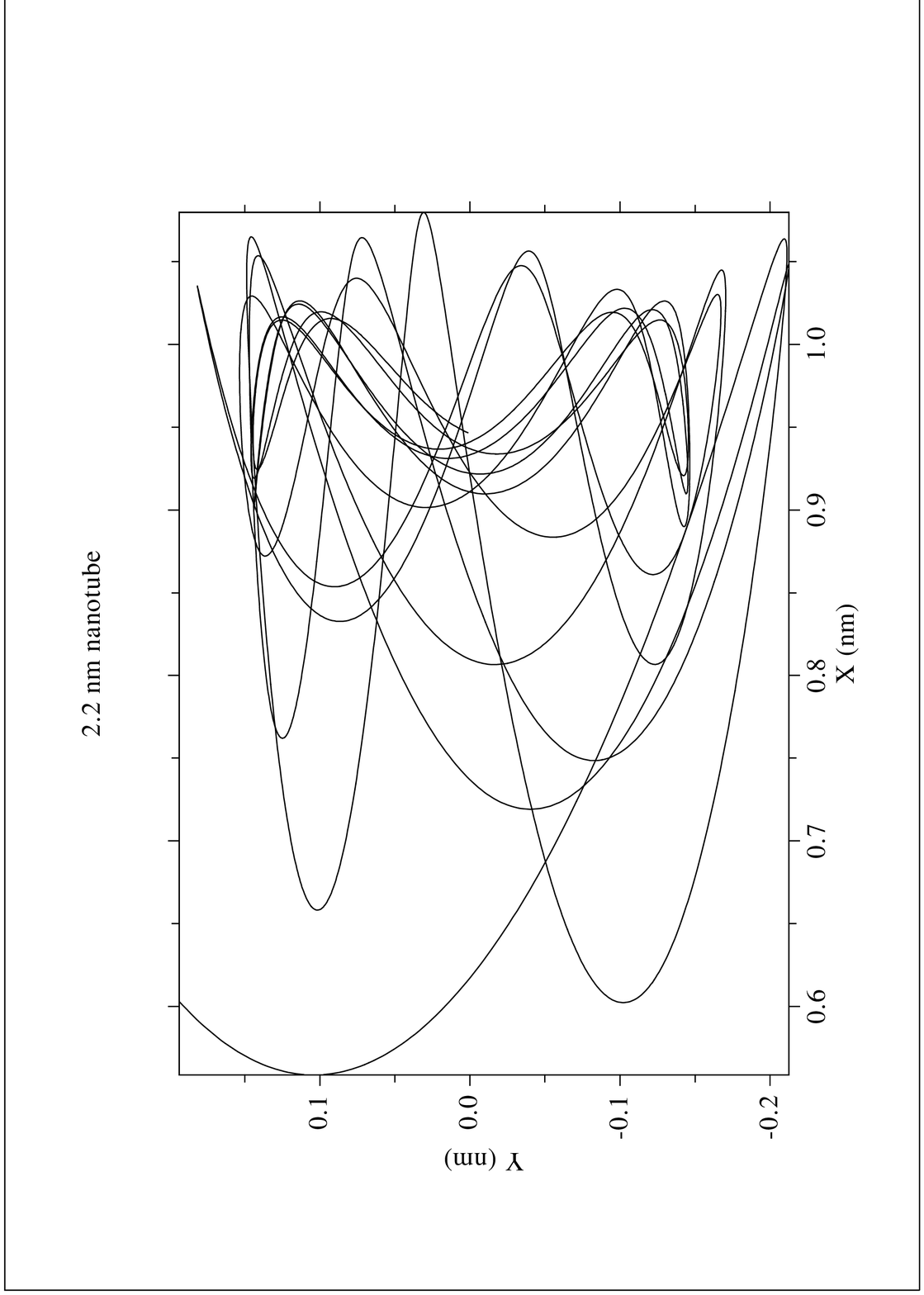,width=9 cm }}
\end{center}
        \caption{(b)
}
        \end{figure}

\pagebreak

\setcounter{figure}{0}
 \begin{figure}[htb]
\begin{center}
\parbox[c]{10.5cm}{\epsfig{file=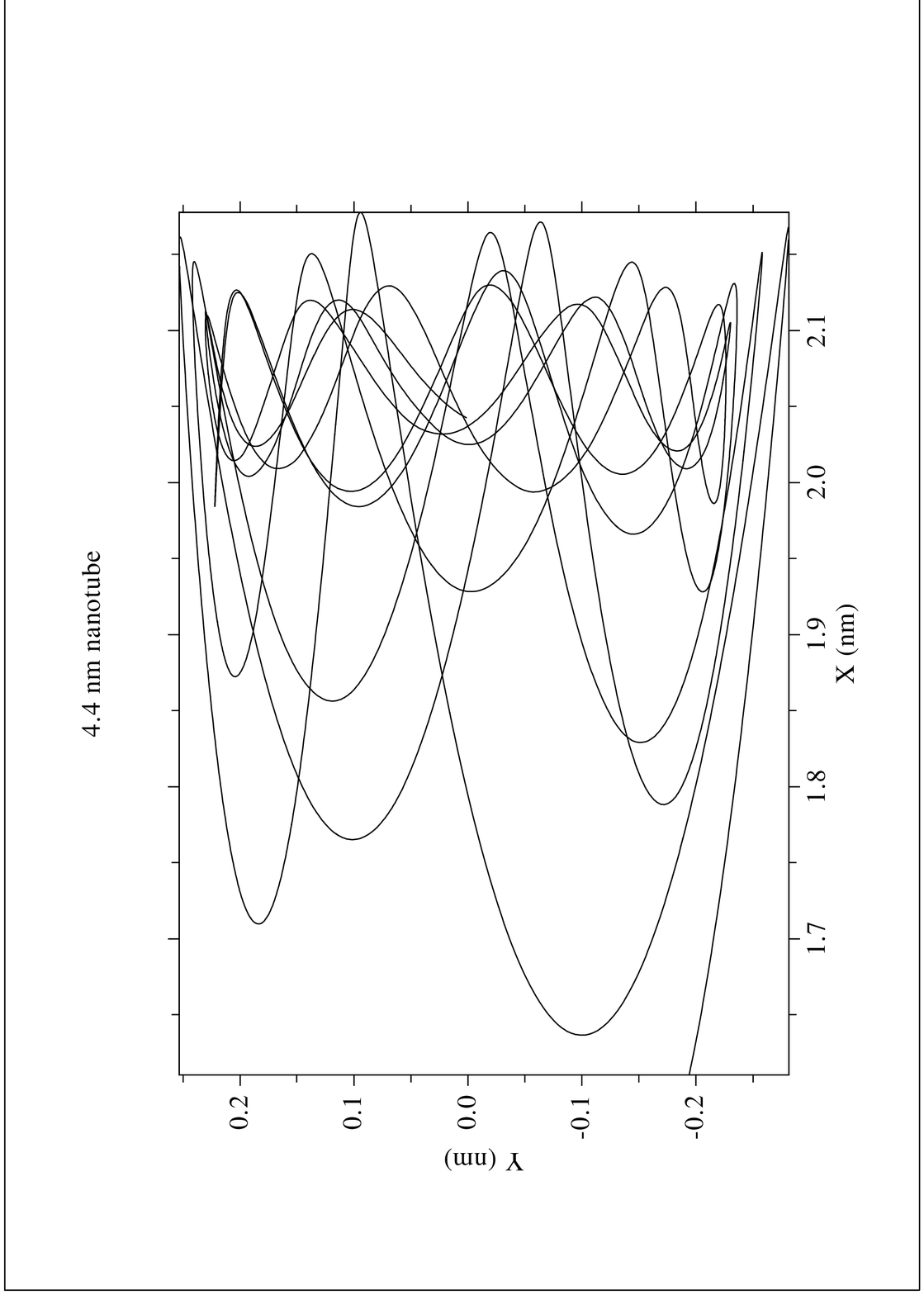,width=9 cm }}
\end{center}
        \caption{(c)
}
        \end{figure}

\pagebreak

\setcounter{figure}{1}
\begin{figure}
	\begin{minipage}{.49\linewidth}
	\begin{center}
  \mbox{\epsfig{figure=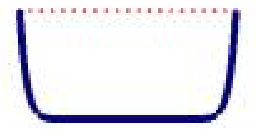,width=.9\linewidth}}
	\end{center}
	\caption{(a)}
	\end{minipage}\hfill
\setcounter{figure}{1}
	\begin{minipage}{.49\linewidth}
	\begin{center}
  \mbox{\epsfig{figure=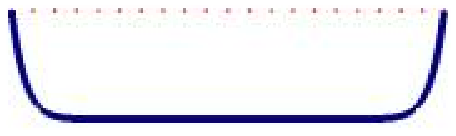,width=.9\linewidth}}
	\end{center}
	\caption{(b)}
	\end{minipage}
\end{figure}
\setcounter{figure}{1}

\begin{figure}
	\begin{minipage}{.49\linewidth}
	\begin{center}
  \mbox{\epsfig{figure=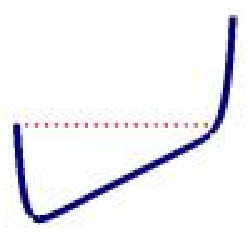,width=.9\linewidth}}
	\end{center}
	\caption{(c)}
	\end{minipage}\hfill
\setcounter{figure}{1}
	\begin{minipage}{.49\linewidth}
	\begin{center}
  \mbox{\epsfig{figure=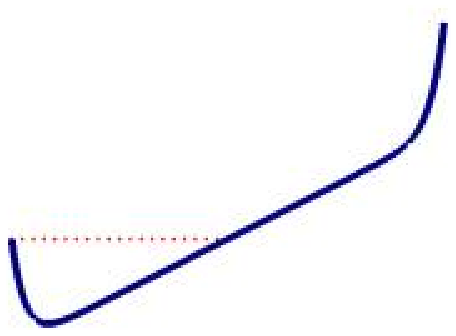,width=.9\linewidth}}
	\end{center}
	\caption{(d)}
	\end{minipage}
\end{figure}
\setcounter{figure}{2}
\pagebreak

 \begin{figure}[htb]
\begin{center}
\parbox[c]{10.5cm}{\epsfig{file=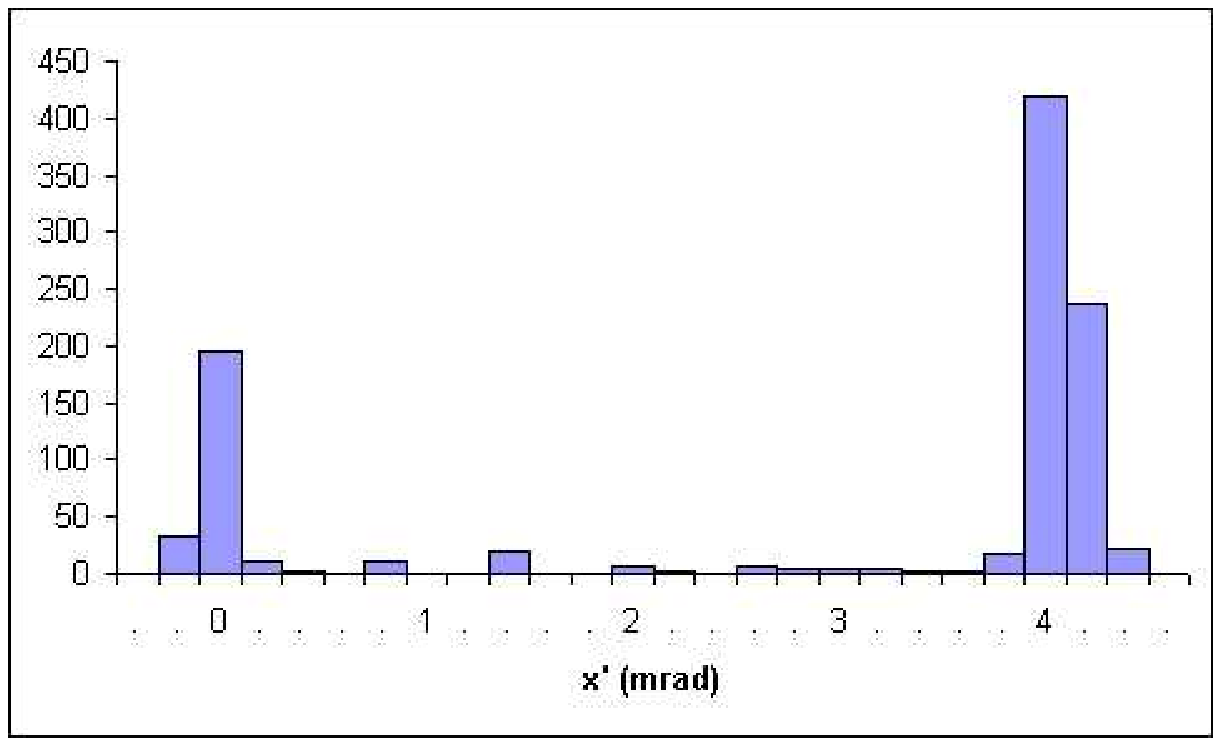,width=9 cm }}
\end{center}
        \caption{
}
        \end{figure}

\pagebreak

 \begin{figure}[htb]
\begin{center}
\parbox[c]{10.5cm}{\epsfig{file=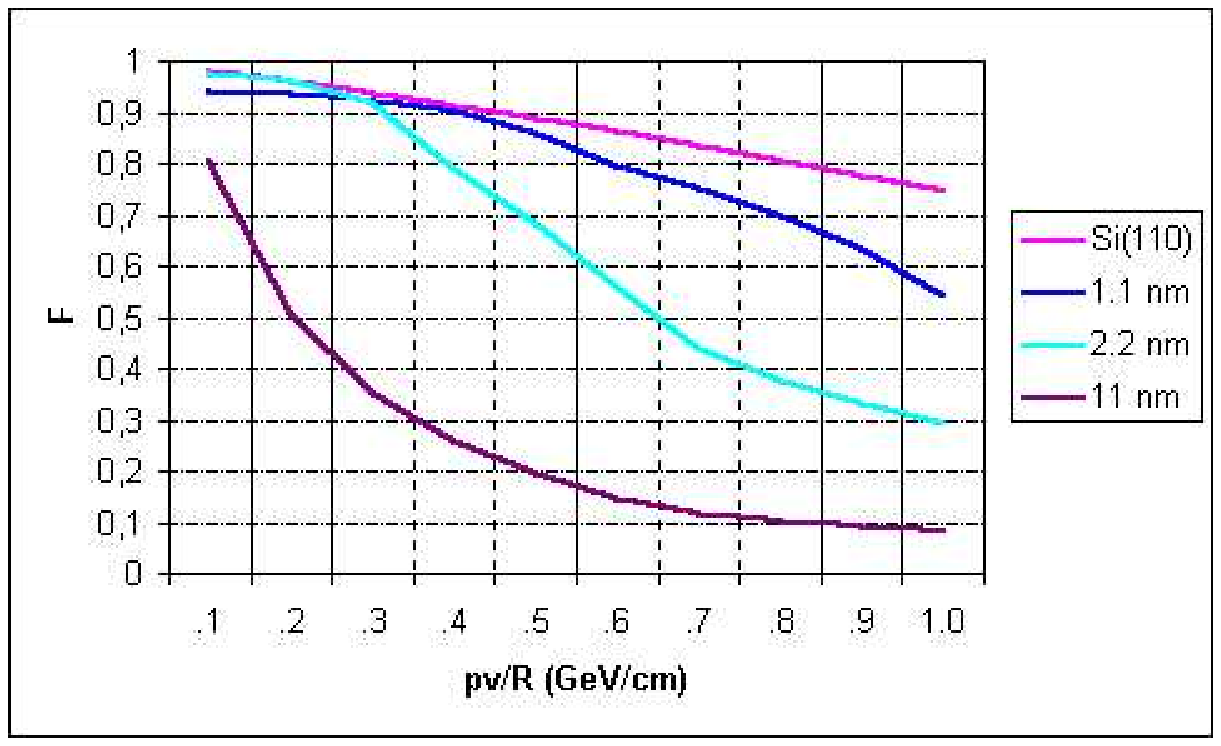,width=9 cm }}
\end{center}
        \caption{
}
        \end{figure}

\end{document}